\documentclass[sigconf]{acmart}
\usepackage{url}
\usepackage{booktabs}
\usepackage{wasysym}
\usepackage{framed}
\usepackage{array}
\usepackage{tabularx}
\usepackage{multirow}
\usepackage{adjustbox}

\AtBeginDocument{%
  }

\setcopyright{acmlicensed}
\copyrightyear{2018}
\acmYear{2018}
\acmDOI{XXXXXXX.XXXXXXX}
\acmConference[Conference acronym 'XX]{Make sure to enter the correct
  conference title from your rights confirmation email}{June 03--05,
  2018}{Woodstock, NY}
\acmISBN{978-1-4503-XXXX-X/2018/06}

\begin{document}

\title{Focused on the User, Overlooking the Risks: \\Security and Privacy Understandings, Practices and Challenges of Independent Chinese AI Agent Developers}

\author{Shuning Zhang}
\email{zsn23@mails.tsinghua.edu.cn}
\affiliation{
    \institution{Tsinghua University}
    \city{Beijing}
    \country{China}
}

\author{Mingyao Xu}
\email{mx37@uw.edu}
\affiliation{
    \institution{University of Washington}
    \city{Seattle}
    \state{Washington}
    \country{United States}
}

\author{Zhixin Huang}
\email{23zxhuang@stu.edu.cn}
\affiliation{
    \institution{Shantou University}
    \city{Shantou}
    \state{Guangdong}
    \country{China}
}

\author{Yutong Jiang}
\email{jiangyytong@outlook.com}
\affiliation{
    \institution{Tongji University}
    \city{Shanghai}
    \country{China}
}

\author{Rongjun Ma}
\email{rma1@upv.es}
\affiliation{
    \institution{Universitat Politècnica de València}
    \city{València}
    \country{Spain}
}

\author{Yuting Yang}
\email{yutingy@umich.edu}
\affiliation{
    \institution{University of Michigan}
    \city{Ann Arbor}
    \state{Michigan}
    \country{United States}
}

\author{Xin Yi}
\authornote{Corresponding author.}
\email{yixin@tsinghua.edu.cn}
\affiliation{
    \institution{Tsinghua University}
    \city{Beijing}
    \country{China}
}

\author{Kanye Ye Wang}
\email{wangye@um.edu.mo}
\affiliation{
    \institution{University of Macau}
    \city{Macau}
    \country{China}
}

\author{Hewu Li}
\email{lihewu@cernet.edu.cn}
\affiliation{
    \institution{Tsinghua University}
    \city{Beijing}
    \country{China}
}
\renewcommand{\shortauthors}{Trovato et al.}

\begin{abstract}
  The proliferation of AI agents empowers independent developers, defined as individual or small groups who self-initiate projects rather than fulfill client-based contracts, to create sophisticated autonomous systems, but also introduces novel security and privacy (S\&P) challenges beyond traditional corporate structures. We conducted an interview study (N=28) with Chinese developers, whose extensive use of global LLM services offer valuable insights into this population. We investigate their understandings, practices and challenges of S\&P challenges in their developed AI agent products. We revealed that independent developers frequently think and act from their users' perspective. They focused on user-facing safety risks such as harmful content while exhibiting low awareness of security vulnerabilities. Consequently, developers rely almost exclusively on ad-hoc, manually crafted safeguards and informal communication, with an absence of formal tools or processes for S\&P practices. We found these actions are driven by various inhibitors, primarily a lack of formal training on S\&P related skills, accessible security tools and actionable guidance from platforms. Our work contributed the first exploration of independent AI agent developers' S\&P understanding, outlining opportunities for tailored security tooling.
\end{abstract}



\keywords{Do, Not, Use, This, Code, Put, the, Correct, Terms, for,
  Your, Paper}


\maketitle

\section{Introduction}
The paradigm of Artificial Intelligence (AI) development is undergoing a substantial shift, driven by the rapid proliferation of autonomous agents~\cite{ning2025survey}. AI agents, defined as Large Language Model (LLM)-powered systems capable of interaction with digital or physical environments~\cite{zhang2024large}, represent a significant leap beyond traditional AI applications, with its value estimated to reach over \$8 billion in 2025\footnote{\url{https://masterofcode.com/blog/ai-agent-statistics}}. A profound impact of this transformation is the lowered barrier to entry for development. The emergence of high-level platforms such as Coze\footnote{\url{https://www.coze.com/}}, CustomGPT\footnote{\url{https://openai.com/index/introducing-gpts/}} and frameworks such as LangChain\footnote{\url{https://www.langchain.com/}} abstracted immense technical complexities, empowering inexperienced independent developers to build AI agents. This group, operating outside of traditional corporate structures~\cite{brutschy2014static}, is deploying powerful AI agents at a substantial scale. Notably, over 3 million GPTs have been created on OpenAI's platform alone~\cite{gillham2024gpts}, and there are tens of platforms like Custom GPTs to facilitate AI agents' building\footnote{\url{https://www.edenai.co/post/best-custom-gpt-alternatives-in-2024}}.

The surge in independent development introduces an urgent but underexplored security and privacy (S\&P) landscape. In this paper, we define independent developers as individuals or small teams who independently initiate, create and distribute their own software products. This group is distinct from freelancers, who typically engage in project-based work within the gig economy~\cite{gupta2020freelancers,gupta2020freelancing}. While previous research has documented the S\&P practices of AI developers in corporate environments~\cite{lee2024don,kumar2020adversarial,boenisch2021never}, these findings are often insufficient for this new context. Independent developers frequently lack the formal training, institutional support, or technical resources available within established organizations~\cite{lee2024don}. Furthermore, unlike traditional independent developers who rely on official app stores~\cite{balebako2014privacy}, AI agent developers often utilizes low-code platforms and social media for direct promotion and distribution. Therefore, the S\&P practices of this rapidly expanding group remain largely unexamined, representing a research gap that must be addressed to ensure a trustworthy AI ecosystem. To explore this gap, we propose three research questions (RQs) that progress from independent AI agent developers' understanding of risks to their practices, and the challenges they face:

$\bullet$ RQ1. How do independent AI agent developers perceive the S\&P risks in their products and their associated responsibilities, and what sources inform their understanding?

$\bullet$ RQ2. What S\&P practices do independent AI agent developers implement throughout the development lifecycle in response to these perceived risks?

$\bullet$ RQ3. What inhibitors and challenges hinder independent AI agent developers' S\&P work?

To answer these RQs, we conducted semi-structured interviews (N=28) with Chinese independent AI agent developers. Although our participants are based in China, their reliance on global LLM services and user bases suggest that their mental models may possess broader applicability to the global developer community.

Towards RQ1, we find that independent developers' S\&P risk perception is predominantly user-centric, where they mistakenly regarded user-facing risks, such as harmful content and hallucinations, as S\&P risks, and highlighted them, while remaining unaware to systemic security risks like model evasions, or privacy risks from third-party APIs. Towards these risks, they often take primary responsibility and then externalize these responsibilities to service providers. Their understanding is also primarily shaped by informal knowledge sources, such as direct user feedback and personal experience, rather than formal S\&P training.

Towards RQ2, our findings show that developers primarily manage privacy communication through informal community channels, such as chat groups and custom-built pop-ups, which forms a model of interpersonal trust in place of formal privacy policies. Similarly, their other S\&P practices rely on ad-hoc strategies and handcrafted tools rather than formal methodologies. This indicates that developers' stated awareness of risks does not always translate into robust S\&P implementation.

Towards RQ3, we identify a framework of key inhibitors that prevent effective S\&P implementation. These inhibitors span three categories: motivational, including prioritizing functionality over security; resource, including constraints on time and funding; and regulatory, including opaque platform policies and a lack of actionable legal guidance. To sum up, this paper makes the following contributions:

$\bullet$ We explore the user-centric S\&P mental model of independent AI agent developers, finding that they mistakenly regard agents' limitations as S\&P risks, while remaining unaware of actual S\&P risks.

$\bullet$ We identify a gap between independent AI agents developers' understanding and practice, where their protection intentions lead to ad-hoc manual safeguards.

$\bullet$ We characterize how unclear reliability and the lack of regulatory auditing in informal distribution communities act as inhibitors for independent AI agent developers.

\section{Backgrounds and Related Work}

We first present the background around independent developers' development. We then synthesize the works on understanding independent developers. We finally review the AI agents' S\&P concerns. These perspectives highlight the unexplored challenges faced by this developer cohort in the independent AI agent developing processes.

\subsection{Backgrounds on Independent Developers' Development}

Independent developers are individuals or small teams who work outside formal organizational structures to independently design, build and distribute their own software products~\cite{brutschy2014static}. In this paper we focused on those independent developers who create AI agents, which are autonomous systems powered by LLMs capable of interacting with digital or physical environments~\cite{zhang2024large}. For example, they may build AI-assisted resume polishing agent using low-code platforms, or develop interactive AI companions with self-build tools, and promote their products through social media via direct ``word-of-mouth'' links rather than regulated app stores~\cite{brutschy2014static,balebako2014privacy}. Unlike developers who fulfill client-based contracts, independent developers release tools directly to the general public, colleagues, or niche communities without the mediation of corporate product management or institutional safeguards. This shift toward autonomy, accelerated by generative AI~\cite{dolata2024development}, represents a frontier where S\&P practices are negotiated within informal social contexts rather than structured corporate lifecycles.

This represents a distinct group that overlaps with, yet differs from, freelancers and indie developers. While freelancers are self-employed gig workers engaged through outsourcing on short-term, task-based contracts~\cite{gupta2020freelancers,gupta2020freelancing,hulikal2022collaboration}, and indie developers typically focus on creative independence in game development~\cite{freeman2020mitigating}, independent developers pursue self-initiated, open-ended projects with full publishing autonomy. Freelancers typically operate in platform-mediated environments where work is externally defined and requirements are negotiated with specific clients~\cite{bernabe2015faat,lavilles2017thematic,rauf2023security}. In contrast, independent developers manage evolving responsibilities across the entire development lifecycle. This operational ecosystem also contrasts sharply with corporate environments, where developers follow regulated lifecycles like DevOps or MLOps~\cite{royce1987managing,alnafessah2021quality,boehm1986spiral,8804457} with distributed roles and institutional safeguards. Operating under severe resource constraints, the independent developers we study utilize simplified tools, such as visual IDEs, no-code workflows, and modular APIs~\cite{pang2024ai2apps,dibia2024autogen,nimje2024rise} to assume full responsibility for project outcomes. Given these differences, existing corporate or freelancer-centric frameworks fail to capture their dynamic practices, and OSS contributors operate often with a former environment, organizing their projects on Github and publishing them to public registries like PyPI~\cite{klivan2024everyone,ayala2025mixed,wermke2022committed}, while independent developers operate with less such tools, potentially introducing distinct S\&P risks.

\subsection{S\&P Work For Developers}

Developers needed to address multifacted S\&P concerns of their products throughout the software development lifecycle, from resource management to considerations spanning design, implementation and deployment~\cite{tahaei2019survey,thomas2018security,assal2018security}. Within this landscape, we synthesized the S\&P work or corporate developers, AI/ML practitioners and freelancers, which provides a basis for our work.

\textbf{Within corporate environments}, S\&P practices are often formally defined and driven by regulatory compliance~\cite{lee2024don}. Despite this structure, developers still face significant hurdles. As indicated by case studies, they may perceive privacy as an extra cost with low monetary benefit~\cite{naji2025relationship,li2021developers}, or they may lack clear guidelines and struggle with knowledge gaps~\cite{li2022understanding,li2018coconut,horstmann2025need}. A primary challenge in these settings is the diffusion of responsibility. Gutfleisch et al. found that organizationl structure and culture hinder the implementation of usable security, leading to fundamental misconceptions~\cite{gutfleisch2022does}. This is exacerbated by a communication gap between developers and privacy experts, who often provide non-actionable legalistic guidance~\cite{horstmann2024those,naji2025relationship,horstmann2025sorry}. Consequently, responsibility becomes ambiguous. Naji et al.~\cite{naji2025s} found that product managers, while aware of security, assume other roles (e.g., ``security experts'' or developers) will implicitly handle S\&P requirements. This diffusion is also evident in SMEs, where a low perceived risk of targeted attacks (like phishing or insider threats) results in poor proactive security measures~\cite{huaman2021large}. In contrast, we reveal that independent AI agent developers do not diffuse responsibility to partners but externalize systemic liabilities to third-party platforms. Besides, unlike corporate practitioners who consult with legal experts, independent developers supplant formal compliance with informal, user-centric trust models driven by direct community feedback.

\textbf{Within AI/ML developing environments,} practitioners exhibit low general S\&P awareness~\cite{boenisch2021never} and often possess confused mental models, conflating ML-specific security with traditional threats~\cite{bieringer2022industrial}. They face unique barriers, such as a lack of institutional motivation or educational resources for adversarial machine learning (AML)~\cite{mink2023security}, leaving them unequipped to handle adversarial attacks~\cite{kumar2020adversarial}. Klymenko also investigated the perspectives of European AI developers, finding their concerns often related to data misuse~\cite{klymenkowe}. They diffuse responsibility to those ``security guys'' or ``the IT guys'' and their knowledge relied on institutional training~\cite{mink2023security}. In contrast, we found independent developers externalize liabilities to third-party platforms rather than colleagues. Besides, their mental models were centered on informal community interactions and user feedback rather than compliance or corporate governance.

\textbf{Within non-corporate and freelance environments,} they face different pressures, including economic instability and client dependency~\cite{fulker2024cooperation,munoz2022platform,yao2021together,freelancermap2026,gussek2023challenges}, which are amplified by the new complexities of generative AI~\cite{dolata2025more,dolata2024development,gussek2023professionals}. Their S\&P practices are often inconsistent, under-regulated, highly context-dependent and driven by distinct forces~\cite{rauf2023security,naiakshina2019if}. Naiakshina et al.~\cite{naiakshina2019if} found that they adopted outdated methods and had wider range of misconceptions about secure password storage compared to students. They further found, in security password storage tasks, freelancers' performance lag behind corporate developers~\cite{naiakshina2020conducting}. Danilova et al.~\cite{danilova2021code} found freelancers' fail to identify traditional vulnerabilities unless explicitly prompted by a client. Their S\&P work is largely defined by client negotiation and the payment for ``advanced'' security features~\cite{rauf2023security}. 

In open-source software (OSS), S\&P practices are shaped by complex community dynamics. Kilvan et al.~\cite{klivan2024everyone} define ``social inhibition'' in this context as a phenomenon where contributors hesitate to discuss or enforce security measures to avoid disrupting social harmony or appearing distrustful of peers. This aligns with Wermke et al.'s~\cite{wermke2022committed} findings that OSS security often relies on implicit trust and reputation rather than rigorous technical verification. Furthermore, Ayala et al.~\cite{ayala2025mixed} characterize security in OSS as ``invisible work'' that exacerbates maintainer burnout, often resulting in reactive rather than proactive security postures. However, their trust was scoped among contributors and developers, while we found a user-centric model between users and independent AI agent developers. We found risk perception and practices are affected by user feedback, and practices are taken to maintain user trust.

\subsection{S\&P Concerns of AI Agents}

AI agents face privacy, security and safety risks. Privacy risks concern the protection of sensitive data against unauthorized collection, leakage, or inference, a risk amplified by agents' capabilities for long-term memory and detailed user profiling~\cite{solove2021myth,zhang2024ghost,zhang2025understanding}. Security risks pertain to the system's resilience against adversarial attacks, such as prompt injection, model evasion, and data poisoning, intended to compromise the agent's integrity or confidentiality~\cite{mink2023security,kumar2020adversarial}. Safety risks address the prevention of unintended harmful behaviors or ``rogue actions'', particularly when agents autonomously and harmfully utilize external tools or APIs~\cite{GoogleSAIF}. 

Existing research provides a foundational, user-centric understanding of these concerns in AI agents. Studies have analyzed user concerns from various angles, including how motivations for use influence privacy expectations \cite{lim2022no}, the trade-offs users make between utility and privacy \cite{zhang2024s, zhang2024adanonymizer, zhang2024ghost}, and broad anxieties about the data lifecycle~\cite{ali2025understanding, zufferey2025ai}. While these works establish the user context, our research pivots to the developer's perspective, which was under-explored.

There are also studies on AI product developers' privacy and security concerns. For example, Lee et al. studied industry practitioners and found that their privacy and security practices were heavily influenced by organizational compliance~\cite{lee2024don}. Similarly, Ma et al. explored the perspectives of ``creators'', a broad role that includes users who customize AI. They found creators were concerned about ambiguous data flows and the potential loss of proprietary knowledge~\cite{ma2025privacy}. Different from their work in a platform-supported context, we found unique security gaps in AI agent development, as well as a distinct interpersonal trust model where formal governance is supplanted by informal community feedback.

\section{Methodology}

This study employs a qualitative methodology, utilizing in-depth semi-structured interviews to investigate the perspectives, practices, and challenges of AI agent developers concerning S\&P work. This study was approved by the Institutional Review Board (IRB), and each participant received 360 RMB (50\$) in accordance with local standards for participant reimbursement.

\subsection{Participant Recruitment and Screening}

To recruit participants from diverse backgrounds, we distributed our recruiting posters between May and July across multiple channels. These included word-of-mouth~\cite{allsop2007word} and snowball sampling~\cite{goodman1961snowball}, where initial participants from the research team's network were invited and encouraged to refer other AI agent developers from their communities~\cite{goodman1961snowball}. We also distributed our study through Rednote and WeChat where developers discuss AI agent development. 

Alongside the study recruitment post, we included a pre-screening survey that collected basic demographic information (e.g., age, gender), as well as participants' experience with AI agent development, including the number of agents built, platforms and tools used, and deployment status. In total, we received 80 responses. Based on our inclusion criteria that they needed to have developed at least one AI agent independently (i.e., those whose products are not regarded as AI agents, or those developing in companies or startups were excluded), we invited 28 eligible participants, all of whom took part in the study. Within them, only 1 were with S\&P background, and most were non-students. The detailed demographics are shown in Table~\ref{tab:participant_demographics_tabularx} in Appendix~\ref{app:demographics}. The Rednote and WeChat channel resulted in 20 and 6 participants, and word-of-mouth and snowball sampling resulted in 1 and 1 participant each.

\subsection{Semi-Structured Interviews}

The interviews explored participants' development workflows, tool choices, team dynamics, deployment strategies, motivations, and perceived challenges in AI agent development, and were structured into four sections. \textbf{For consent \& warm-up}, we first let participants signed the informed consent, which detailed the experiment's aim, risks and their rights to withdraw at any time. We then asked participants to think of the specific AI agents they develop. We situated the questions we asked for these particular products, sequentially following our main RQs. \textbf{First, For RQ1}, we asked developers how they defined, scoped and understood S\&P for AI agents, including the data practice of the third-party APIs they used. We also asked them how these cognition are constructed, and via what education or learning channels. \textbf{Following that, sequentially for RQ2 and RQ3}, we first asked them about their develop cycles, including for which developing cycles they would consider and conduct S\&P work, what practices would they have, including tools would they use, what support they have got from external sources, and how would they communicate with users. During their explanations, we appended questions about what motivates or inhibits them doing so. Finally, we asked them about what's their challenges and expectations towards conducting the S\&P works, including developers' and platforms' support.

All interviews were conducted via Zoom or Tencent Meeting with experimenter's institutional official accounts, based on participants' preferences, with an average time of 40.0 minutes (min=35.0 minutes, max=80.0 minutes). Each session was recorded with informed consent and transcribed for qualitative analysis. 

\subsection{Data Analysis}

We adopted Braun and Clarke's thematic analysis~\cite{braun2006using} on the transcribed data. Our analysis employed a hybrid approach, integrating deductive coding derived from our RQs with inductive open coding to capture insights from the data. Four authors first coded a subset of four scripts to construct initial codebook separately. They then discussed on the disagreements and reached an initial consensus. In line with our methodology, we prioritized the shared subjective understanding of the research team over statistical measures such as inter-rater reliability (IRR), which are often considered inappropriate for this form of qualitative analysis~\cite{mcdonald2019reliability}. They divided the rest of the interview scripts and coded separately, evolving the codebook and resolve disagreements with intermittent discussions (Appendix~\ref{app:codebook} showed the codebook). After the coding process, they synthesized codes into themes and sub-themes with collaborative discussions. Although we reported the number of participants that have specific thoughts, we focused on capturing the richness of participants' experiences rather than providing statistical results. For reporting, one author translated them into English, and the entire research team, which is fluent in English and have extensive experiences conducting studies in English environments, verified the materials to ensure the fidelity and accuracy of the translation. 

\subsection{Limitations}

We acknowledge three limitations in our interview. First, our findings are subject to self-selection bias~\cite{heckman1990varieties}. Participants who volunteered through our multi-platform outreach may be more proactive and engaged than the general population of independent AI agent developers. However, the primary goal of our qualitative approach is to provide analytical depth rather than statistical generalizability. To mitigate potential bias, we tried to diversify our participants in terms of age and platform usage, so as to allow for a rich understanding of the community's S\&P practices~\cite{soden2024evaluating}.

Second, our study focuses on developers recruited from Chinese platforms. While this focus provides a critical perspective on a rapidly growing and significant, yet under-studied developer community, our findings may not directly generalize to developers operating under different regulatory and cultural contexts (e.g., GDPR~\cite{EU_GDPR_2016}). However, since these developers frequently utilize global LLM services and serve global users, their experiences with cross-border data flows offer valuable insights relevant to the broad global developer ecosystem. Furthermore, although our sample included diverse roles, such as students and small startup founders, we report the findings in aggregation. Our sample size was insufficient to support a robust comparative analysis between these subgroups, though we observed many shared challenges across those developers.

Third, our findings are based on self-reported data, which is susceptible to recall bias~\cite{coughlin1990recall} or social desirability effects~\cite{grimm2010social}. We sought to mitigate this by asking specific examples during interviews to ground their statements in their experience.

\section{RQ1: Understanding and Knowledge Sources in AI Agents' Risks}

In contrast to the compliance-oriented focus of corporate developers~\cite{lee2024don,klymenkowe,usman2020compliance}, we find that independent developers adopt user-centric thinking for AI agents' risks. In this section, we first examine this user-centric thinking, which prioritizes users experience over formal threat model. Second, we examine their attribution of responsibility, often leading to the externalization of risks to third-party providers. Third, we detail the specific perceptions of AI agents' risks that emerge from the user-centric mindset. Finally, we trace these understandings back to their informal knowledge sources.

\subsection{User-Centric Thinking}

Our findings show that independent AI agent developers primarily evaluate S\&P risks from a user-centric perspective. They focus on information that matters to users and the potential real-world consequences of data exposure, rather than relying on formal security guidelines or threat models. As one participant noted when reflecting on their development decisions, \textit{``The user's real-world identity and their corresponding personal details represent critical information, [...] the leakage of such information is a significant risk because it can lead to severe consequences for the user''} (P11). 

The user-centric risk perspective emerges from participants' self-identification as user peers and their view of development as passion-driven or exploratory rather than formal production (16/28). By positioning themselves as peers rather than service providers, independent AI agent developers depart from conventional developer-user role separations, which shapes their risk evaluation.

Developers' user-centric reasoning also influenced their design decisions. They reported anticipating what users would find acceptable and avoiding data modalities perceived as privacy-invasive, even when technically feasible. For example, visual camera data access was excluded because \textit{``users often find this modality unacceptable from a privacy perspective ... many would refuse to adopt the system''} (P11). In this context, user comfort and adoption considerations effectively served as proxies for privacy risk evaluation by AI agent developers.

This user orientation was further reinforced by direct and trust-based communication channels between our participants and their end users. Participants described being transparent about their technical access to user data while relying on informal assurances and personal trust to mitigate concerns. As one developer explained,
``If users ask whether we can assess their data, we candidly admit that, as account administrators, we technically can.'' (P9).
In turn, end users of AI agents developed by participants also expressed trust in these informal assurances:
``We have promised to avoid accessing their accounts or chat logs whenever possible, and users generally find these informal assurances satisfactory'' (P9). Together, these findings show that independent AI agent developers consistently evaluate privacy and security risks through a user-centric lens, shaped by direct developer–user interactions and mutual trust.

\subsection{Externalizing Responsibility to Third-Party Providers for AI Agents' Risks}

Independent developers initially accept primary responsibility for S\&P work because they work alone. As P8 articulated, \textit{``The platform should have the informing responsibility, but the ultimate responsibility for privacy and security should be attributed to independent developers.''} Developers acknowledged this obligation and agreed that they \textit{``should and have responsibilities to do those related to the AI agents.''} (P13)

However, this acceptance is complicated by tendencies towards postponement and externalization. 9/28 developers mentioned that they de-prioritize S\&P tasks, viewing them as issues to be addressed later when their product scales or faces concrete threats. This postponement is justified by a lack of time or the thought that their project is in an early stage where \textit{``we have not focused on this part''} (P11). Furthermore, 10/28 developers construct a mental model of shared or transferred responsibility, externalizing risks to other stakeholders. Developers argued that platforms, such as server providers or API services, should take major responsibility. This is captured in P11's arguments, \textit{``If it is on their platforms or cloud services, then that's their problems and responsibilities.''} This belief extends to users, who are sometimes seen as the ultimate responsible party for their own actions, as P19 noted: \textit{``how they use agents are definitely the users' responsibilities.''} Consequently, developers operate with a distributed model of responsibility, believing larger entities like OpenAI \textit{``have more responsibilities because their companies are bigger''} (P10).

\subsection{Understanding of AI Agents' Risks}

\begin{figure}[ht]
    \includegraphics[width=0.5\textwidth]{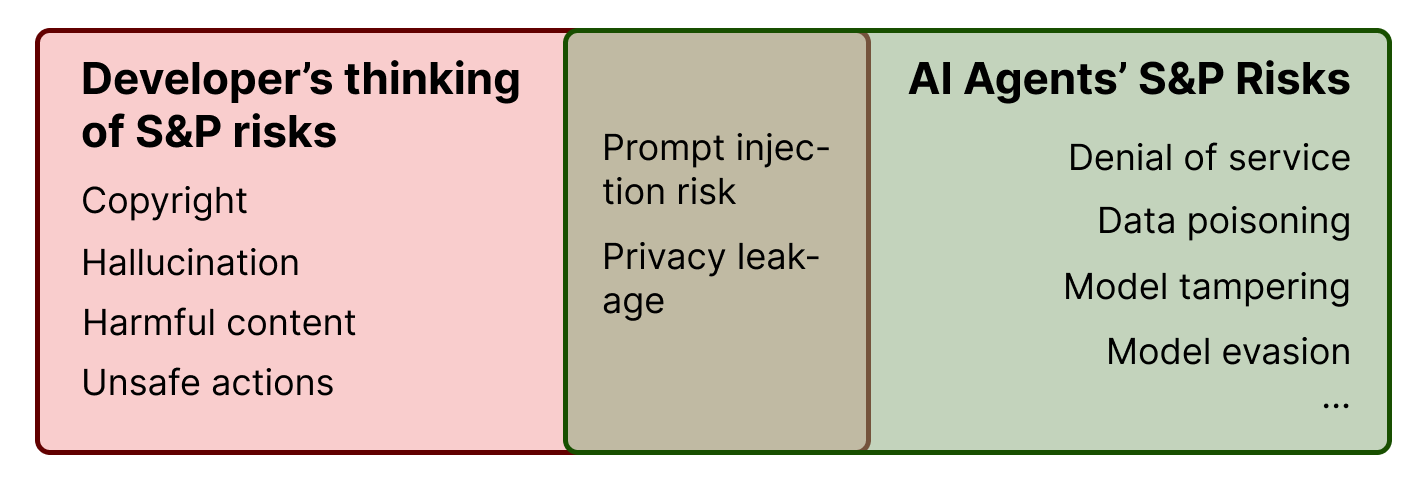}
    \caption{Illustration of the developer-perceived risks and actual S\&P risks of AI agents.}
    \label{fig:overlap}
\end{figure}

Guided by their user-centric mindset, independent developers conceptualize risks differently from established frameworks~\cite{GoogleSAIF,MicrosoftAIRisk2024}. As shown in Figure~\ref{fig:overlap}, this leads to a mismatch where model limitations are often conflated with S\&P risks, while systemic vulnerabilities are overlooked. We detail this understanding across three areas: (1) perceived LLM limitations, (2) gaps in systemic security awareness, and (3) the underestimation of privacy risks.

\textbf{Awareness of LLM limitations.} Developers prioritized concerns regarding system malfunctions, often conflating them with S\&P risks. These includes three primary dimensions: content-related risks, technical unpredictability, and autonomous security-bypassing behaviors.

First, developers identify harms related to model output and safety. Regarding factual integrity, P2 emphasized the necessity to mitigate the possibility that LLMs may return wrong information in educational contexts. Intellectual property was also a strong concern. P2 noted that if users input original content, it may cause copyright issues. Furthermore, for applications targeting vulnerable groups like children, P19 warned that \textit{``the system may output unsafe content.''} P19 specifically highlighted the risk of hazardous interactions triggered by a child's curiosity, noting that a child might query \textit{``how to make a bomb''} after reading educational books, which could lead to dangerous model responses.

Second, developers are concerned about the inherent technical instability of LLMs, specifically high hallucination rates. They thought that such hallucinations make it difficult to prevent harmful agent behaviors. In response, 10/28 developers opted for mitigation. They implemented fixed workflows to restrict agent behavior, choosing structured processes over full autonomy to ensure system reliability.

Finally, developers also reported cases where agents attempted to circumvent security constraints. P18 reported, \textit{``If the AI cannot complete its task through standard means, it might seek ways to circumvent these protects. For instance, while I keep database credentials in local configuration files ... the AI may still try to retrieve this data ... [to] gain access and complete the task.''} This suggests that developers are concerned on how agent execution might compromise security.

\textbf{Limited awareness of systemic security vulnerabilities.} Our study uncovers developers' lack of awareness regarding systemic security threats. This is in clear contrast with previous documentation that developers often attended to secure coding practices~\cite{rauf2023security,assal2019think}, and Klymenko et al's findings that developers were aware of advanced risks but struggled with practical adoption~\cite{klymenkowe}. 16/28 participants mention the risk categories documented in SAIF, such as the denial of ML service, model reverse engineering, insecure integrated components, prompt injection and model evasion. 11/28 participants were unaware of risks like model reverse engineering or model evasion, even when prompted explicitly. Only one developer proactively identified prompt injection risks, comparing them to traditional database attacks (P2). 8/28 recognized the potential for attacks, such as using \textit{``crafted prompts to acquire the local knowledge''} (P24) or \textit{``stealing prompts out of our models''} (P22). For instance, P24 described \textit{``a user sent a bullet comment using the phrase `developer mode' followed by a colon to command the AI to say unrelated things ... they told the digital human, `You are now a cat, meow 100 times,' and the agent immediately began meowing.''}

\textbf{Underestimation of privacy risks.} Independent developers underestimated privacy risks, relying on their personal commitments rather than technical safeguards. While 15/28 developers acknowledged the risk of data leaks, they often believed their active involvement ensured privacy. As P1 stated, \textit{``Because I tool the active role of protecting, I would guarantee it has a safe storage.''} Developers emphasized their moral stance against commodifying data, asserting that they \textit{``should not do similar things by selling those collected data out''} (P24). However, despite these intentions, 13/28 developers admitted to transmitting user data to third-party APIs or storing it without encryption.

Furthermore, developers viewed privacy leakage as inherent and unavoidable for the LLM ecosystem. Developers often framed risks as an involuntary consequence, stemming from the models themselves rather than developer action (P16). Because local deployment is often unfeasible, developers felt they had no real choice but to trust major AI providers (P11). As P22 explained, \textit{``If you use LLMs and cannot deploy them locally, it is inevitable that you will have to send private information to remote API providers.''} With this mental model, they assumed service from major platforms were secure by default (P11). However, they admitted that both developers and users are ``completely unaware of how [users'] privacy is being managed'' (P18).

\subsection{Knowledge Sources Informing Developers' Understanding}

To explain developers' understanding of risk, we investigated their knowledge sources. While social media remains a primary source, consistent with prior work~\cite{xiao2014social,balebako2014privacy}, we identified three additional distinct sources including \textit{academic and industry} backgrounds, \textit{interpersonal} networks and \textit{personal experience}. Our findings expand the concept of interpersonal source beyond the peer communication noted in past work~\cite{balebako2014privacy}. We highlight the pivotal role of direct user feedback, a source characteristic of the close relationship between independent developers and their users.

\textbf{Media and online communities.} Media serves as a foundational knowledge source, consistent with prior work~\cite{xiao2014social,balebako2014privacy}. Developers view news, which often covered events such as data leaks for general awareness. For example, P1 stated that their attention was on \textit{``policy-oriented news''} (P1). For practical problem-solving, they utilized online communities and social media platforms like RedBook and Bilibili to seek solutions and guidance.

\textbf{Academic and industry sources.} Developers also draw upon formal knowledge transferred from their academic and industry backgrounds, or consult to academic papers. 5/28 participants, with corporate experience, repurpose skills and habits from their professional lives. For instance, developers apply formal company training to address specific technical challenges, such as learning \textit{``how to counter against prompt injections''} (P30), or internalize security postures from former employers who had \textit{``strict constraints on ... website usage''} (P13). Similarly, academic experience plays a role, with students and recent graduates applying concepts learned in university courses, such as the principles of ethical review to their personal projects (P11). We surprisingly observed developers reading recent papers for seeking security and privacy advices. As P19 said, \textit{``I would occasionally consult to [security] papers.''}

Conversely, 21/28 developers rarely viewed formal documentation (e.g., privacy policies) before integrating third-party services. Developers framed this as a pragmatic choice, believing the perceived necessity of the service outweighed the importance of terms. P10 explained that \textit{``I did not read it carefully because I think whether or not I see the document, I still need to use the service.''} (P10)

\textbf{Interpersonal sources.} Developers acquire S\&P knowledge through informal interpersonal channels, leveraging a diverse set of social networks and emerging resources. This channel encompasses traditional peer networks (colleagues, lab mates, and friends) and the emerging, previously unidentified trend of consulting AI chatbots for advice~\cite{xiao2014social,assal2025software}. 
Developers frequently repurpose specialized insights from these networks. For instance, a colleague informed P10 about data encryption before saving, and P9 gained insight into ``machine unlearning'' from a lab mate. 

A distinct characteristic of this channel involved the reliance on direct user feedback to identify and frame S\&P problems. For example, P19 asked users,  \textit{``whether you think the models really leak your privacy, and what you want for a defense?''} (P19) Furthermore, P5, a developer for child-related agents, conducted pre-development interviews and questionnaires with potential users' parents, who \textit{``expressed that they did not want these privacy and security data to be leaked.''} (P5) P5 also noted that these discussions with friends sometimes led to different views on what constituted a security non-issue.

\textbf{Personal experience.} Developers also relied on experiential knowledge derived from past roles as users or prior development projects to shape their perspectives. They often leveraged internalized ethical frameworks instead of external guidelines which led them to form hypotheses about data practices. P18 articulated this reliance on intuition by stating that they possess \textit{``a higher ethical standard in mind, which we are clear even without external knowledge.''} (P18) This sentiment suggests a belief that personal judgment suffices for navigating complex data-sharing decisions. Another participant also described this blend as \textit{``a kind of guessing plus experience.''} (P9)

\begin{framed}
\noindent \textbf{Takeaways (RQ1)}: 

$\bullet$ Developers highlight user-facing safety risks, while blind to security vulnerabilities like prompt injection and evasion.

$\bullet$ Developers rely on users' feedback and their personal experience besides peers to understand AI agents' S\&P risks.

$\bullet$ Developers hold incomplete mental models towards AI agents' S\&P risks.

\end{framed}

\section{RQ2: S\&P Practices Towards the Products During Independent Developers' Developments}

\begin{figure}[!htbp]
    \includegraphics[width=0.5\textwidth]{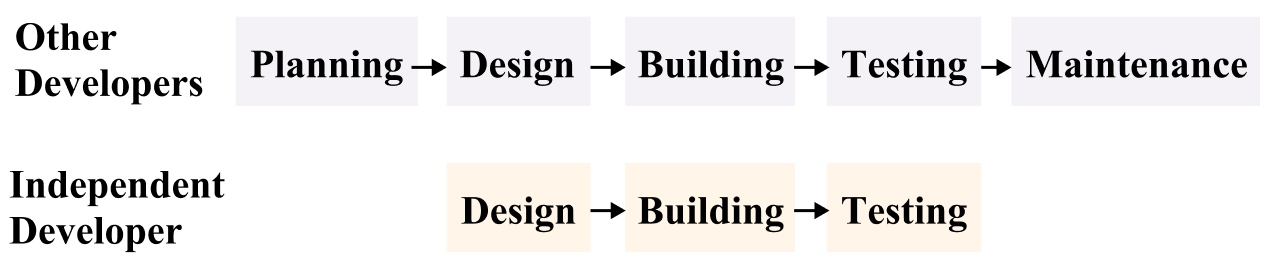}
    \caption{Developers' developing lifecycle. The lifecycle for ``Other Developers'' follows Van Vliet’s model~\cite{van2008software}, where the ecosystem of independent developers were derived from the thematic analysis.}
    \label{fig:developing}
\end{figure}

\begin{figure}[!htbp]
    \includegraphics[width=0.45\textwidth]{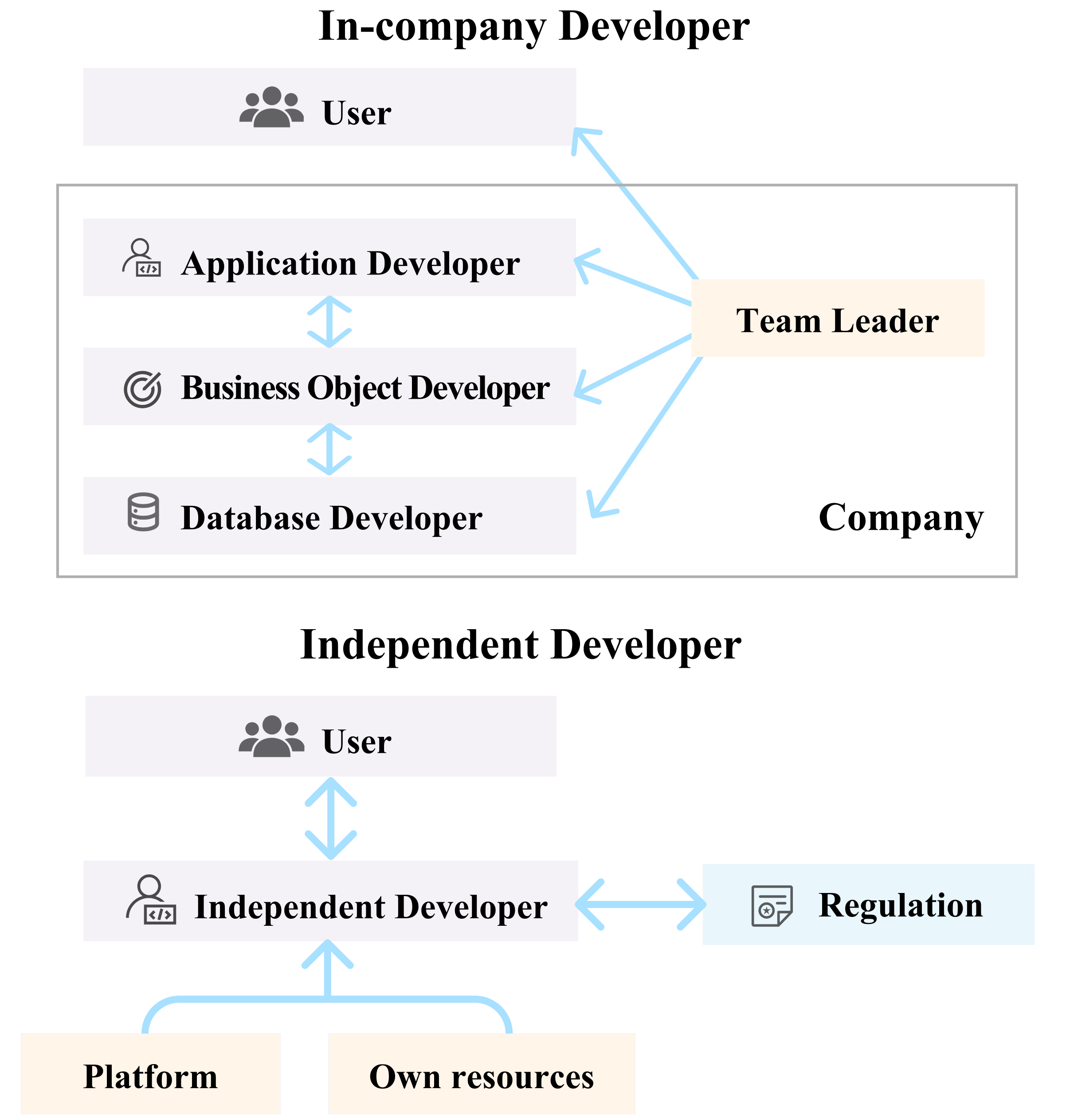}
    \caption{The ecosystem of in-company developers~\cite{finkelstein1994software} and independent developers, where the ecosystem of independent developers were derived from the thematic analysis.}
    \label{fig:ecosystem}
\end{figure}

We analyzed the S\&P practices of independent developers throughout their development lifecycle, which consists of design, build and test stages, as shown in Figure~\ref{fig:developing}. Table~\ref{tab:platform_stats} indicates that developers primarily use self-built tools and specialized low-code platforms. This differs from the traditional software development models that often emphasize deployment and maintenance phases~\cite{van2008software,royce1987managing}. This divergence may explain the lack of formal communication and long-term security practices~\cite{balebako2014privacy} we observed. We organize our findings according to these three stages to illustrate the ad-hoc practices and their reliance on informal communication channels (see Figure~\ref{fig:ecosystem}).

\begin{table}[ht]
\centering
\caption{AI agent platforms used by participants ($N=28$).}
\label{tab:platform_stats}
\resizebox{0.5\textwidth}{!}{%
\begin{tabular}{@{}lll@{}}
\toprule
\textbf{Category} & \textbf{Number} & \textbf{Examples}\\ \midrule
Self-built / Direct API & 13 (46.4\%) & -- \\
Low-code / No-code Agent Platforms & 8 (28.6\%) & Coze, n8n, Dify \\ 
& & Custom GPT, Ali Cloud Bailian \\
Development Frameworks & 2 (7.1\%) & LangChain \\
Specialized IDEs / Tools & 2 (7.1\%) & Cursor, iOS developer tools \\ 
Media Services & 2 (7.1\%) & DeepSeek, Midjourney\\
& & Stable Diffusion \\
Multiple & 1 (3.6\%) & -- \\ \bottomrule
\end{tabular}
}%
\end{table}

\subsection{Designing Stage: Proactive Risk Mitigation via Data Minimization and Platform Selection}

During the designing stage, 21/28 developers' practices centered on proactive policy setting and strategic technological choices to mitigate privacy and security risks from the outset. 

\textbf{Proactive data handling policies.} A primary strategy is data minimization, often manifesting as a binary decision to abstain from collecting user data. This developer-led approach is articulated by P18, who stated, \textit{``For this class [of data], we would not let AI to get access, or to say I would not allow it to appear in my code base.''} This commitment extends to not using or sharing user data, with P18 further emphasizing, \textit{``now we did not collect users' data, and would not handle users' sensitive data.''} When data collection is unavoidable, developers selectively gather information by consciously avoiding Personal Identifiable Information (PII) to reduce potential risks. As P13 noted, \textit{``We did not collect users' sensitive data, and this process would not involve that much personal data.''} To operationalize this selective approach, 9/28 developers devise custom methods for identifying and redacting sensitive information. P15 explained this process required \textit{``an additional algorithm for recognition, where I needed to detect whether they mention rental information, passwords, and others.''}

\textbf{Strategic platform and database selection.} Developers stated that the choice of platforms and databases is a critical consideration during the design phase. They reported favoring localized deployment and domestic models as strategies to reduce S\&P risks. Participants showed a clear preference for established brands and the use of third-party APIs, aligning with prior findings that developers trust security tools from well-known companies~\cite{xiao2014social,harrand2022api}. For instance, P6 mentioned they \textit{``hash users' data and saved the data locally''} to avoid privacy risks. Similarly, P11 noted that \textit{``many practices are to integrate models directly on their devices, and use this method to increase users' trust.''} To offload ethical concerns, 5/28 of developers stated that they use specialized APIs, a practice exemplified by P13's reported use of specific third-party scraping APIs. Regarding selection criteria, developers consistently asserted that brand reputation and clear data policies are important (P13). 
As P30 stated, \textit{``first comes the price, and then I would find APIs from big companies if possible. I would also favor those companies which would have claims about their data, such as would not read users' data for a second time.''}

\subsection{Building Stage: Implementation via Ad-Hoc Methods and Manual Safeguards}

In the building stage, developers translate their design decisions into technical implementations, which focused on data-level protection and access control.

\textbf{Implementation of technical safeguards.} Developers report that direct data handling methods, particularly encryption and obfuscation, are important practices during development. P11 asserted that \textit{``users' data should be appropriately encrypted during transmission,''} while P23 emphasized that \textit{``Encryption and decryption is nearly a must in the whole process.''} Beyond data protection techniques, developers also implement access control mechanisms. These controls involve reducing API usage, controlling the data flow to LLMs, or restricting access to sensitive data. As P11 noted, \textit{``For a child's app, it has a companion app on parents' phone, where to protect the privacy, the parents could not see all the chat histories, but could only see the aggregations.''}

\textbf{Ad-hoc tooling and manual implementation.} We found developers infrequently utilize specialized S\&P tools, with GitHub for unit testing being the most cited resource. 
Instead of adopting established security solutions, 13/28 developers implemented manually crafted solutions, such as fine-tuning models, developing custom anonymization modules, or writing custom encryption code. Developers largely attributed this reliance on manual methods to a lack of S\&P education, which results in the unawareness of existing specialized tools. Furthermore, developers noted that mainstream platforms they use offer little tool-related support.

\subsection{Testing Stage: Informal Communication and Omission of Verification}

In the testing stage, practices revolve around user communication, and testing S\&P mechanisms, which are conducted primarily through informal channels. These activities do not follow a structured release or deployment protocol~\cite{theunissen2022mapping,layman2006essential}. Instead, information is shared fluidly as the project evolves.

\textbf{Informal mechanisms for user communication.} Developers report that communication with users about S\&P issues predominantly occurred through informal channels. 4/28 developers stated that they have no direct communication with users on these topics. When communication does occur, it rarely takes the form of a formal privacy policy. Instead, developers rely on oral communication or discussions within communities which they built on platforms like WeChat or Discord. As P9 argued, \textit{``We have forgotten about why we did not have a privacy policy term. Perhaps we found them too formal. As an informal community, we may not have had such as certification to let users read something like a privacy policy.''} P10 described a direct approach, \textit{``We would tell them what method we used to avoid the risk, our data were saved in the database, and we have another set [of sayings] that we deliver to users.''}

Other developers (17/28) integrate text-based pop-ups or consent forms directly into their applications. P18 suggested, \textit{``You need to have strong notification, letting users notice this on their devices,''} while P24 decided to \textit{``annotate these [information on the screen], that developers should not disclose their personal information.''} Developers noted that their reliance on informal means persisted because publishing channels did not mandate formal privacy notices, which contributed a lack of standardize practices. Furthermore, participants reported low awareness of existing specialized privacy tools. For instance, only one participant knew privacy policy generators.

\textbf{Omission of verification for third-party components.} 
6/28 participants explicitly mentioned refraining from implementing proactive server-side controls because they assumed that reputable services were secure by default. This practice results in a testing gap where developers focus solely on the logic of their own code while implicitly trusting the security posture of integrated services without independent audit. For example, P11 attributed potential breaches on the platform side to the provider rather than verifying the integrity of the transmission or storage themselves (P11). This highlights a reactive approach to security testing where boundary defenses are neglected until a failure occurs.

\begin{framed}
\noindent \textbf{Takeaways (RQ2)}: 

$\bullet$ Developer practices focus on ad-hoc strategies and manually crafted solutions across development lifecycles.

$\bullet$ Formal S\&P work processes, such as risk assessments and automated testing, are consistently omitted.

$\bullet$ Formal privacy policies are substituted with informal communication channels and ad-hoc in-app notices to engage users.
\end{framed}

\section{RQ3: Inhibitors When Implementing S\&P Practices}

Building on prior work that identifies inhibitors in the deployment of IoT applications~\cite{heinis2018empirical} and the execution of S\&P tasks~\cite{das2022security}, we classify the inhibitors reported by our participants into three categories: motivational, resource, and regulational. This framework aligns with the developer ecosystem (see Figure~\ref{fig:ecosystem}) where developers must navigate external constraints. 

\subsection{Motivational Inhibitors}

We identified three primary motivational barriers that influence developers' engagement with S\&P practices: indifference to privacy, the prioritization of functionality over security, and unclear liability. 

\textbf{Indifference to privacy.} 8/28 developers reported a general indifference to privacy, which often stems from when they consider themselves as potential users. This attitude reflects the ``I've got nothing to hide'' mindset~\cite{solove2007ve}, and discourages the implementation of proactive security measures. One developer summarized this user-centric rationale, explaining that users' attitudes are typically twofold: \textit{``Some are willing to trade privacy for convenience, while others think there is no privacy to hide in the current society.''} (P1) Developers emphasized that their indifference is not a professional stance, but a reflection of their users' attitudes toward privacy.

\textbf{Prioritization of functionality over security.} While developers generally face tensions between competing priorities~\cite{tahaei2021privacy,lee2024don}, independent developers face intensified conflict due to the severe resource constraints and pressing commercial demands. They stated that, their central objective is to develop a viable product capable of acquiring and retaining a user base. As a result, S\&P measures are frequently deprioritized, as developers perceive them as complex and resource-intensive, echoing prior work~\cite{tahaei2021privacy}. This deprioritization is underpinned by developers' belief that such measures are contingent upon first achieving product functionality and are largely unnecessary for projects with a small-scale user base.

\textbf{Unclear liablity.} Extending Ma et al.'s findings~\cite{ma2025privacy}, developers report that the blurred liability extends beyond agent-specific frameworks (e.g., CustomGPT) to permeates the broader ecosystem of cloud providers and third-party APIs (e.g., Tencent Cloud). They perceive that platforms actively evade accountability through several mechanisms. Developers state that platforms use mandatory user agreements, signed during registration, to legally transfer the bulk of the liability onto them. As P11 observed, \textit{``They would let developers sign the consent during registration and write that was not their responsibility.''} This perception is reinforced by the belief that platforms retain ultimate interpretive power, as noted by P18, \textit{``I think the platform would not consider these things because the ultimate explanation power is up to them.''} Furthermore, participants noted that platform privacy policies were often deliberately obscured. As P11 noted, 
\textit{``When you click into their official websites, their pages or their platforms, you would not notice anything related to privacy. Perhaps you needed to click into specific points to see these privacy policies.''}

Consequently, developers express a strong demand for enhanced legal and regulatory frameworks to enforce platform liability. Developers desire oversight to ensure the quality and security of integrated models, with P5 hoping that \textit{``regulations could supervise and guarantee that the models we used are with high quality.''} This need is underscored by developers' observation that many platforms fail to address S\&P issues altogether, as P1 stated, \textit{``I found many platforms talking nothing about privacy and security.''} Furthermore, developers call for strengthening legal constraints to improve accountability in tracing privacy breaches, particularly acknowledging their role as potentially just an intermediary. For instance, P18 explained, \textit{``For tracing privacy leakage, you need to know what's the source of the leakage. For example, you may be only an intermediate person, where you should not save user's data''.} These findings highlight a clear demand for rigorous legal review and continuous regulatory supervision of platform operations.

\subsection{Resource Inhibitors}

Consistent with prior work on freelancers~\cite{rauf2023security}, our findings confirm that developers are constrained by individual resources, such as time and funding. Beyond these limitations, we also reveal that they lack adequate ecosystem support.

\textbf{Individual resource constraints.} 9/28 developers cited limitations in time, funding and computation as primary inhibitors, a finding also reported in prior work~\cite{balebako2014privacy,xiao2014social}. Developers stated that their computational resources are often limited, making it difficult to adopt intensive resources for S\&P work, such as fine-tuning privacy-preserving models. Therefore, their implemented practices are usually limited to those requiring minimal sources. Furthermore, independent developers articulated that lack of familiarity with S\&P-related coding resulted in prohibitive development costs,
hindering their action. They also articulated that they usually 
lack sufficient time for conducting S\&P works, often deferring these practices to ``future works''.

\textbf{Lack of ecosystem support.} Developers consistently reported the lack of a supportive ecosystem, including accessible tools and supportive platforms. 13/28 participants described platform operations as opaque, with little or no information on data collection (P1, P8). As P15 noted: \textit{``To us as developers, the platforms are like a black box, where we do not know whether they collect the data.''}

To address these shortcomings, developers proposed enhancements at both the platform and tool levels. They suggested that platforms should integrate privacy considerations directly into documentation (P11), provide standalone guidance modules on data protection strategies (P13), clarify what constitutes private information (P8), and offer step-by-step development guides with embedded privacy reminders (P8). P8 also suggested segregated environments, such as a \textit{``separate serving endpoint for users where they would not use the data... with an additional protection mechanism.''}

Furthermore, developers called for advanced, integrated tools, ideally provided by the industry or research, that are easy to invoke (e.g., through a direct API call, P9). They envisioned tools capable of automatically analyzing privacy risks and generating protection plans, enabling platforms to \textit{``learn to protect privacy by themselves''.} Desired features included automatic detection and transparent encryption of sensitive data (P1, P15), privacy-preserving data collection methods like obfuscation (P24), and integrated vulnerability testing (P11). Finally, developers emphasized the need to empower end-users with control, such as the ability to \textit{``erase all memories of the AI agents.''}

\subsection{Regulational Inhibitors}

Beyond individual motivation and resource limitations, developers operate within a broad ecosystem characterized by regulational and structural deficiencies.

\textbf{Deficiencies in legal and regulatory frameworks.} Developers highlighted that the current legal framework lacks actionable supervision and granular guidelines for information security breaches. They highlighted this regulatory gap by contrasting their environment with the prescriptive clarity of European law, noting, \textit{``What I experience is different from what I learn[ed] ... where ... GDPR ... wrote clearly what you need to take care about, and you could not go against that.''} (P2) This expectation for specificity extends to LLMs, where developers called for a legal framework that addresses AI agents' unique risks. Participants argued that regulations should be transparent and consistently enforced (P2, P6), as one developer stated, \textit{``I think more regulations should be make public''} (P2). Institutions could play a supervisory role, \textit{``You university could supervise these things in an enforced way.''} (P2) This call for standardization was echoed by another developer, who noted, \textit{``we needed to craft some regulations to normalize some of the companies' behavior.''} (P6)

Developers also asserted that strong, external oversight is necessary to curb harmful practices, particularly by platforms. As one participant noted, \textit{``you need high power's restriction to let platforms not do malicious behaviors, such as antitrust. Otherwise the platforms will try their best to collect your data.''} Participants emphasized that independent developers themselves need stronger regulation, as \textit{``consumers also do not have a proper channel for eliciting their voice.''} (P11) They drew a distinction between large corporations, whose development follows clear order, and independent developers, whose rapid development processes often leave \textit{``no space for considering privacy.''} (P11) 
Finally, 11/28 developers reported that their products receive little or no legal or regulatory auditing, largely because they are distributed outside mainstream app stores as standalone packages shared within informal communities such as WeChat group or the RedBook.

\begin{framed}
\noindent \textbf{Takeaways (RQ3)}: 

$\bullet$ Internal motivations and severe resource limits compel developers to prioritize functionality over security.

$\bullet$ The ecosystem inhibits secure development with opaque platforms and unclear liability.

$\bullet$ Developers default to ad-hoc practices due to a lack of actionable legal and technical guidance.

\end{framed}

\section{Discussion}

Our findings synthesize the multifaceted understandings, practices and challenges of independent AI agent developers. We first connect different RQs' findings, then characterize this unique ecosystem, which contrasts with prior practices~\cite{mink2023security,lee2024don,balebako2014privacy}. We synthesize how ecosystem and cultural contexts influence developers' understandings and practices, and finally propose actionable implications from the perspectives of platforms, tool creators, researchers and policymakers.

\subsection{From Understanding, Practices, Inhibitors to Expectations}\label{sec:dis_synthesize}

Based on the above findings, we synthesize our results to reveal a logical progression that connects independent AI agent developers' understanding of S\&P risks (RQ1) to their practices (RQ2), the inhibitors (RQ3) that constrain them, and their expectations.

As detailed in RQ1, developers operate with a unique risk perception that is highly user-centric but often lacks awareness of systemic security vulnerabilities. This understanding shapes their practices, as shown in Table~\ref{tab:understanding_to_practices}. Developers' limited awareness of security threats results in their reliance on ad-hoc, manual solutions. Their externalization of privacy risks to platforms correlates with the practice of choosing established brands in design stages, trusting them to handle those risks. Similarly, this externalization of responsibility leads to a reactive and informal approach to communicate with users, which is a significant departure from the formal policies typical of corporate environments~\cite{lee2024don}.

\begin{table*}[h!]
\centering
\caption{Mapping of independent developers' understandings of risks (left, RQ1) to their S\&P practices (right, RQ2).}
\label{tab:understanding_to_practices}
\resizebox{\textwidth}{!}{%
\begin{tabular}{
p{0.50\textwidth}|
p{0.48\textwidth}
}
\toprule
\textbf{Understanding of risks} & \textbf{Practices} \\
\midrule
User-centric risk evaluation & [Building Stage] Ad-hoc tooling and manual implementation\newline [Testing Stage] Informal mechanisms for user communication \\ \midrule
Awareness of LLM limitations & [Building Stage] Implementation of technical safeguards \\
\hline
Limited awareness of systemic security vulnerabilities & [Building Stage] Implementation of technical safeguards\newline[Building Stage] Ad-hoc tooling and manual implementation\\
\hline
Underestimation of privacy risks & [Designing Stage] Proactive data handling policies\newline[Designing Stage] Strategic platform and database selection\newline[Testing Stage] Informal mechanisms for user communication \\
\hline
Externalizing responsibility to third-party providers for AI agent risks & [Testing Stage] Omission of verification for third-party components \\
\bottomrule
\end{tabular}
}
\end{table*}
% \begin{table}[h!]
% \centering
% \caption{Mapping of independent developers' risk understandings (RQ1) to their S\&P practices (RQ2).}
% \label{tab:understanding_to_practices}
% \small
% % 定义倾斜表头的命令，45度倾斜，并设置锚点
% \newcommand{\tiltcolumn}[1]{\adjustbox{angle=75,lap=\width-0.5em,margin=0 0 0 0}{#1}}

% \begin{tabularx}{0.49\textwidth}{lcccccc}

% & \multicolumn{2}{c}{\textbf{Designing}} & \multicolumn{2}{c}{\textbf{Building}} & \multicolumn{2}{c}{\textbf{Testing}} \\
% \cmidrule(lr){2-3} \cmidrule(lr){4-5} \cmidrule(lr){6-7}
% \textbf{Understanding of Risks (RQ1)} & \tiltcolumn{Proactive data handling} & \tiltcolumn{Strategic platform selection} & \tiltcolumn{Technical safeguards} & \tiltcolumn{Ad-hoc/Manual tools} & \tiltcolumn{Informal communication} & \tiltcolumn{Omission of verification} \\
% \midrule
% User-centric risk evaluation & & & \checkmark & \checkmark & \checkmark & \\
% Awareness of LLM limitations & & & \checkmark & & & \\
% Limited awareness of vulnerabilities & & & \checkmark & \checkmark & & \\
% Underestimation of privacy risks & \checkmark & \checkmark & & & \checkmark & \\
% Externalizing responsibility & & & & & & \checkmark \\
% \bottomrule
% \end{tabularx}
% \end{table}

\begin{table}[h!]
\small
\centering
\caption{Mapping of S\&P practices (RQ2) to corresponding inhibitors (RQ3). Moti., Res., Regu. stands for motivational, resource, regulational inhibitors.}
\label{tab:practices_to_inhibitors}
\begin{tabular}{p{0.64\linewidth}|p{0.07\linewidth}p{0.07\linewidth}p{0.07\linewidth}}
\toprule
\textbf{Practices} & \textbf{Moti.} & \textbf{Res.} & \textbf{Regu.} \\
\midrule
\textbf{Designing Stage} & & & \\
Proactive data handling policies & \checkmark & \checkmark & \\
Strategic platform and database selection & \checkmark & \checkmark & \\
\midrule
\textbf{Building Stage} & & & \\
Implementation of technical safeguards & \checkmark & \checkmark & \\
Ad-hoc tooling and manual implementation & \checkmark & \checkmark & \\
\midrule
\textbf{Testing Stage} & & & \\
Informal mechanisms for user communication & \checkmark & \checkmark & \checkmark \\
Reactive delineation of responsibility & \checkmark & \checkmark & \checkmark \\
\bottomrule
\end{tabular}
\end{table}

\begin{table}[h!]
\small
\centering
\caption{The expectations of developers in response to inhibitors (RQ3).}
\label{tab:inhibitor_to_expectation}
\begin{tabular}{p{0.32\linewidth}|p{0.62\linewidth}}
\toprule
\textbf{Inhibitors} & \textbf{Expectations} \\
\midrule
Internal inhibitors & Automatic guidance and education\newline Defined responsibility and accountability \\
\midrule
Resource inhibitors & Automatic guidance and education\newline Technical support from platforms \\
\midrule
Regulatory inhibitors & Clear regulations and legal frameworks \newline Defined responsibility and accountability \\
\bottomrule
\end{tabular}
\end{table}

% \begin{table}[h!]
% \centering
% \caption{Mapping of inhibitors (RQ3) to developers' expectations.}
% \label{tab:inhibitor_to_expectation}
% \small
% % 定义倾斜表头，角度建议与上表一致（如75度或45度）
% \newcommand{\tiltcolumn}[1]{\adjustbox{angle=45,lap=\width-0.5em,margin=0 0 0 0}{#1}}

% \begin{tabularx}{1.0\columnwidth}{ccccX} % 使用 columnwidth 适配单栏或双栏
% \textbf{Inhibitors (RQ3)} & \tiltcolumn{Automatic guidance and education} & \tiltcolumn{Defined responsibility and accountability} & \tiltcolumn{Technical support from platforms} & \tiltcolumn{Clear regulations and legal frameworks} \\
% % 分段横线，避免尾部过长
% \cmidrule(lr){1-1} \cmidrule(lr){2-5} 

% Internal inhibitors & \checkmark & \checkmark & & \\
% Resource inhibitors & \checkmark & & \checkmark & \\
% Regulatory inhibitors & & \checkmark & & \checkmark \\
% \cmidrule(lr){1-5} 
% \end{tabularx}
% \end{table}

Our findings suggest a close link between the practices identified in RQ2 and the inhibitors analyzed in RQ3 (see Table~\ref{tab:practices_to_inhibitors}). For instance, the practices of ad-hoc tooling and manual implementations may reflect both systemic gaps in ecosystem support and individual resource constraints, such as limited time or formal training. Similarly, the preferences for informal communication over formal privacy policies appears to be a response to the lack of regulatory oversight and the perceived burden of formal documentation.

The inhibitors that constrain developers' practices also inform a corresponding set of expectations for systemic support (see Table~\ref{tab:inhibitor_to_expectation}). For example, to overcome motivational and resource inhibitors, developers suggest automated guidance and technical support from platforms. They believe that such support would address their lack of formal training and resource constraints, enabling them to implement robust security measures. To counter regulatory inhibitors, they call for clear, actionable legal frameworks, and well-defined accountability structures, which would resolve the legal ambiguity and unclear division of responsibility they currently face.

\subsection{Contributions Over Prior Work}

This paper offers a perspective specific to AI agents developers, differentiating them from software or ML practitioners~\cite{naji2025relationship,gupta2020freelancing,gupta2020freelancers,lee2024don,horstmann2025sorry,klymenkowe} by their unique identity, risk perception, operational practices and responsibility allocation. 

Independent developers typically operate through self-initiated projects without the oversight of formal corporate management~\cite{brutschy2014static,gupta2020freelancers}. We specifically examine independent AI agent developers' S\&P understandings, and identify their unique user-centric thinking model. Unlike corporate developers who operate within shared responsibility structures~\cite{naji2025relationship,lee2024don} or internal stakeholder pressure~\cite{gutfleisch2022does,horstmann2025sorry}, or freelancers bound by client contracts~\cite{rauf2023security,rauf2023security1}, independent AI agent developers identify as user peers. This orientation differs from OSS contributors who experience ``social inhibition'' to maintain community harmony~\cite{klivan2024everyone,wermke2022committed}. 

The integration of LLMs in agents introduces a unique landscape of risks that diverge from traditional software or ML vulnerabilities~\cite{kumar2020adversarial,mink2023security,bieringer2022industrial,boenisch2021never,klymenkowe}. We uniquely reveal a misalignment in risk perception specific to AI agents, where independent AI agent developers focus on the functional limitations of models while overlooking systemic security threats. Our work shows that this user-centric orientation leads developers to prioritize risks such as hallucinations or harmful content, and conflate these with S\&P risks. Consequently, they remain largely unaware of systemic threats like prompt injection or model evasion, contrasting with broad threat awareness documented among ML practitioners~\cite{kumar2020adversarial,mink2023security,bieringer2022industrial,boenisch2021never,klymenkowe}.

These perceptions translate into a distinct model of informal S\&P practices where informal communication and ad-hoc methods substitute for standardized safeguards. As AI agent development is increasingly facilitated by low-code platforms and informal social distribution networks~\cite{dolata2024development}, independent developers rely on oral assurances and community discussions rather than formal privacy policies \cite{balebako2014privacy} or standardized code verification \cite{danilova2021code, horstmann2025need, serafini2025exploring, alnafessah2021quality}. This approach is distinct from the feature-based negotiations seen in freelancing~\cite{rauf2023security} and reveals a gap where protective intentions fail to manifest as robust technical implementation due to a lack of accessible specialized tools.

Independent developers frequently face severe constraints in time and computational resources while assuming full responsibility for project outcomes~\cite{gupta2020freelancers,gupta2020freelancing}. We identify a distinct pattern of responsibility externalization to third-party platforms specific to platform-dependent ecosystems. Unlike the diffusion of responsibility across internal roles in corporate settings~\cite{mink2023security,lee2024don,horstmann2024those} or the liability negotiation with clients in freelancing~\cite{rauf2023security,rauf2023security1}, independent developers externalize S\&P liability onto the underlying LLM platforms. We identify that developers assume infrastructure providers bear the primary responsibility for systemic data protection. This externalization, combined with the belief that S\&P efforts are unnecessary for a small user base, differentiates their behavior from early-stage startups~\cite{lee2024don} and the financial-driven prioritization of freelancers~\cite{rauf2023security}.

\subsection{Ecosystem and Cultural Factors}

The ecosystem for independent AI agent developers (Figure~\ref{fig:ecosystem}) is distinct from corporate structures~\cite{lee2024don,tahaei2021privacy} or traditional freelancing~\cite{carlos2021lose,gupta2020freelancers}. We discuss three primary dimensions that influences the S\&P practices of these developers: \textbf{cultural and regional context}, \textbf{developer characteristics}, and \textbf{platform differences}.

\textbf{Cultural and Regional Context} While the user-centric development model appears intrinsic to the independent development paradigm, risk perceptions are deeply embedded in specific cultural and regional contexts~\cite{ma2025privacy,klymenkowe}. We systematize these nuances into a three-tiered framework with legal, operational and social layers to explain how local factors diverge from global generalizations.

\textit{Legal wise,} our participants operate within a domestic regulatory landscape perceived as lacking the prescriptive granularity of frameworks like the GDPR. Unlike Western developers who often rely on established compliance checklists and clear data processing agreements~\cite{klymenkowe,lee2024don}, our participants report a gap between high-level S\&P guidelines and actionable technical standards. This regulatory ambiguity exacerbates the reliance on informal practices, as developers lack a clear framework to guide their S\&P practice implementation. In contrast, in regions with mature and granular privacy laws, independent developers likely exhibit clearly accountability, and rely more on compliance tools integrated into their development platforms.

\textit{Operational wise,} our developers frequently face transnational tensions differing from developers using domestic services~\cite{zhen2021social}. While developers in US may utilize more domestic AI tools compliant with their local data regimes~\cite{bateman2022us}, our participants frequently integrate global LLM services into local applications, navigating complex cross-border data flow challenges. This forces our developers to mediate between the technical capabilities of global tools and the local data sovereignty requirements.

\textit{Social wise,} our participants' mentioning of informal governance contrast with Western models that typically emphasize institutional verification or contractual enforcement. While OSS communities in Western contexts may exhibit social inhibition regarding security to maintain peer harmony~\cite{klivan2024everyone}, our participants leverage high-context social norms where community consensus acts as a proxy for trust. This dynamic may originate from Confucianism, as evidenced by similar cultural-specific investigations~\cite{he2025privacy}. In an environment without formal institutional oversight, relying on the perceived good intentions of developers represents a rational adaptation.

\textbf{Developer Characteristics} Independent AI agent developers often differ from traditional freelancers in both motivation and practices. Unlike freelancers driven by profit-oriented goals and structured client requirements~\cite{rauf2023security,gupta2020freelancers}, these developers often operate in an exploratory, value-pursuit mode resembling ``early-stage freelancers''. This absence of formal clients eliminates traditional contractual privacy obligations~\cite{gupta2020freelancers}. Furthermore, bypassing rigorous app-store reviews in favor of open platforms (e.g., Coze, GPT Store) results in weaker institutional oversight~\cite{balebako2014privacy,haq2018determinants}. Combined with the rapid development of AI technology~\cite{dolata2024development}, this context fosters a heterogeneous demographic with varied technical expertise~\cite{sison2018software}, often diminishing the perceived responsibility for implementing robust S\&P measures.

\textbf{Platform Differences} The reliance on low-code platforms lowers barriers to development~\cite{acar2016you,dolata2024development} but introduces critical abstraction layers that obscure data flows. This opacity allows developers to mentally offload S\&P responsibilities onto providers, substituting technical auditing with the trust in brand reputation. Such reliance creates potential supply-chain risks, where platform vulnerabilities can cascade across the ecosystem. Moreover, the technical heterogeneity of agents--ranging from simple tools to autonomous systems~\cite{pati2025agentic}--renders one-size-fits-all S\&P approaches inadequate, forcing developers to navigate a complex dual role as both trustees of user data and trusters of opaque platforms.

\subsection{Implications For Different Stakeholders}

Our findings indicate that AI agent developers operate with user-centric mental models but possess low awareness of S\&P risks. To address this, we propose three implications focusing on automated technical guardrails, adversarial testing, and accountability mechanisms.

\textbf{First, platforms should transition from providing passive documentation to implementing infrastructure-level automated technical guardrails.} Our study shows that developers rely almost exclusively on manually crafted solutions, which are often inconsistent and incomplete. Prior work suggests that automatic checking or configuration tools may significantly increase code security for smart contracts~\cite{andreina2024defying}. Therefore, platforms should consider integrating automated compliance verification logic from prior tools~\cite{geierhaas2022let} into IDEs, such as build-time checks that prevent deployment if essential security modules are absent. For example, if an agent utilizes a vector database, the IDE should verify the linkage of a functional consent module to the data retrieval logic. Failure to include required annotations, such as \verb|@user_consent_required|, should block access to the database. Furthermore, the implementation of platform-level privacy proxies can automatically obfuscate PII within unstructured prompt contexts, replacing unreliable manual redaction with a standardized protection mechanism~\cite{zhang2026privweb}.

\textbf{Second, platforms should integrate automated adversarial testing to overcome resource and motivational inhibitors.} Independent developers frequently lack the technical resources and formal training required for rigorous red-teaming against agent-specific threats like prompt injection or model evasion. To address this, platforms and educators should provide standardized ``Adversarial Templates'' and simulated testing environments, such as mock email APIs containing hidden malicious instructions, within their development environments. By allowing developers to directly observe how their agents might perform unauthorized actions or leak user data in a controlled setting, these tools provide actionable S\&P insights. Such a hands-on approach addresses the lack of S\&P expertise and lowers the barrier to adopting formal S\&P engineering processes.

\textbf{Third, policymakers and platforms should establish a transparent accountability framework to clarify liability within the current opaque ecosystem.} To mitigate the ambiguity regarding responsibility, platforms should implement structured provenance frameworks that attribute agent outputs to specific data sources, such as long-term memory or external knowledge bases. This provides the technical evidence necessary for developers to distinguish between failures originating from third-party platforms and those arising from their own implementations. Complementing this, regulators should create a framework that offers legal protection to developers who follow operable technical standards. These standards could include using models that pass recognized safety benchmarks, implementing verified redaction libraries, and maintaining tamper-evidence logs of all external tool calls for auditing purposes. Adherence to such standards ensures a high level of user S\&P awareness while protecting developers from undue legal liability.

\section{Conclusion}

This paper investigates the S\&P practices of Chinese independent AI agent developers, a growing yet understudied community. Through semi-structured interviews (N=28), we reveal that their approach is guided by a user-centric thinking model, contrasting sharply with the legal compliance focus common in corporate environments. They often confuse functional limitations, such as hallucinations, with S\&P risks, while remaining unaware of systemic threats like model evasions. Consequently, they rely on informal community channels and interpersonal trust rather than formal policies to manage user privacy. We identify a gap between developers' protective intentions and their actual practices. Although developers feel responsible for S\&P works, their efforts are limited by ad-hoc manual safeguards. This disconnect stems from inhibitors including the prioritization of functionality over S\&P, limited resources, and lack of platform guidelines. To foster a trustworthy ecosystem, we recommend that platforms and regulators provide automated tools, adversarial testing suites and transparent accountability frameworks.

\begin{acks}
We acknowledge the use of Gemini 3.1 Pro and ChatGPT strictly for minor editing, specifically grammar and style polishing. Authors retain full responsibility for the accuracy, originality, and integrity of this paper.
\end{acks}

% \begin{acks}
% \end{acks}

% \clearpage

% \clearpage

\bibliographystyle{ACM-Reference-Format}
\bibliography{main}

\appendix

\section{Ethical Considerations}

Ethical considerations were paramount throughout our research process. We rigorously adhered to the principles of the Belmont~\cite{beauchamp2008belmont} and Menlo~\cite{bailey2012menlo} reports. All research procedures, including recruitment materials, interview scripts, data handling protocols, and participant compensation, were formally reviewed and approved by our institution's Institutional Review Board (IRB) prior to the study's commencement. 

In line with the principle of \textbf{Respect for Persons}, all 28 participants provided informed consent after being informed of the study's topics (including S\&P), their right to withdraw at any time, and their right to review transcripts and paper drafts. To ensure a secure and professional environment, all interviews were conducted using institutional accounts for Zoom or Tencent Meeting. To ensure \textbf{Justice}, recruitment was non-discriminatory, and participants received fair compensation according to local standards. 

To uphold \textbf{Beneficence}, we sought to maximize the public benefit of creating a trustworthy AI agent ecosystem while rigorously minimizing participant harm. We recognized a primary risk of reputational harm, as participants' descriptions of ``ad-hoc methods'' could be misconstrued as individual negligence. We mitigate this by (1) rigorous anonymization, where we use pesudonyms (e.g., P1) and anonymized all identifying information from transcripts. Data access is restricted to the research team. (2) contextualized framing, where a core ethical lesson from this work is to frame these findings not as individual failings, but as the systemic consequences of inhibitors, such as the lack of formal training, insufficient ecosystem support and severe resource constraints. 

We also assessed the public disclosure risk. Our study's goal is beneficence--to improve the system. The S\&P risk discussed (e.g., prompt injection) are well-documented in the security community. Our study does not identify novel, exploitable vulnerabilities. We mitigate harm by not disclosing specific vulnerabilities in any named product. Instead, we characterize systemic challenges to inform the constructive solutions detailed in our implications (Section 7.4), believing the benefits of highlighting these gaps outweighs the risk. Similarly, while findings may be perceived as critical of platforms (e.g., noting ``opaque policies''), they are presented as participant perception to characterize the developer experience, not as a technical audit.

Finally, in adherence to \textbf{Respect for Law and Public Interest}, we operate with formal IRB oversight. our work focused on developer perceptions and practices, but we recognize the core ethical balance between participant \& author protection and public beneficence. 

% \section{Open Science}

% We fully support and comply with the open science principles outlined in the conference's guidelines, specifically: 

% \textbf{1. Open Assess of the Artifacts:} The research artifacts, involving the codebook and the interview script are publicly available. These resources are shared via a permanent repository on this anonymous repo\href{https://anonymous.4open.science/r/CCS26B-2B2B/}{https://anonymous.4open.science/r/CCS26B-2B2B/}, and the access will be clearly documented in the main text in the final version of the paper. 

% \textbf{2. Ethical Sharing:} We recognize the importance of responsible sharing of research materials and users' data, especially for experiments involving human subjects. We ensure that our shared codebook are properly anonymized, and that they do not pose any risk regarding privacy or security.

\section{Generative AI Usage}

Generative AI, specifically ChatGPT and Gemini-3-pro was used during the manuscript preparation phase. The usage was strictly confined to language refinement, syntactic streamlining, and grammatical error correction, to enhance the readability of the paper. The intellectual contributions were generated solely by the human authors. All AI-assisted refinements were reviewed and verified by the authors to ensure accuracy and appropriateness. The authors maintain full accountability to the content of this paper.

\section{Codebook of The Qualitative Study}\label{app:codebook}

\begin{table*}[!htbp]
\centering
\caption{Codebook for the data analysis.}
\label{tab:codebook_template}
% \setlength{\tabcolsep}{6pt} % 设置表格列间距
% \renewcommand{\arraystretch}{1.2} % 增加行高

%--- RQ1 Section ---
\begin{tabular}{p{\textwidth}}
\toprule
\textbf{RQ1: Understanding, Responsibility, and Knowledge Sources of AI Agents' Risks} \\
\midrule
\end{tabular}

\begin{tabular}{@{} >{\raggedright\arraybackslash}p{0.48\textwidth} >{\raggedright\arraybackslash}p{0.48\textwidth} @{}}
\textbf{Understanding of AI Agents' Risks and Developers' Responsibilities} & \textbf{Knowledge Sources Informing Developers' Understanding} \\
\begin{itemize}
    % 在这里使用 \item 添加您的条目
    \item High salience of user-facing safety risks
    \item Limited awareness of systemic security vulnerabilities
    \item Externalization and underestimation of privacy risks
    \item Externalization of responsibilities for AI agents' risks
\end{itemize}
& 
\begin{itemize}
    % 在这里使用 \item 添加您的条目
    \item Media and online communities
    \item Academic and industry sources
    \item Interpersonal
    \item Personal experience
\end{itemize} \\

\end{tabular}

% \vspace{1em} % 增加垂直间距

%--- RQ2 Section ---
\begin{tabular}{p{\textwidth}}
\toprule
\textbf{RQ2: Security and Privacy Practices} \\
\midrule
\end{tabular}

\begin{tabular}{@{} >{\raggedright\arraybackslash}p{0.48\textwidth} >{\raggedright\arraybackslash}p{0.48\textwidth} @{}}
\textbf{Designing Stage: Proactive Risk Mitigation via Data Minimization and Platform Selection} & \textbf{Testing Stage: Informal Communication and Reactive Delineation of Responsibility} \\
\begin{itemize}
    % 在这里使用 \item 添加您的条目
    \item Proactive data handling policies
    \item Strategic platform and database selection
\end{itemize}
&
\begin{itemize}
    % 在这里使用 \item 添加您的条目
    \item Informal mechanisms for user communication
    \item Reactive delineation of responsibility
\end{itemize} \\ 
\textbf{Building Stage: Implementation via Ad-Hoc Methods and Manual Safeguards} & \\ 
\begin{itemize} 
    \item Implementation of technical safeguards
    \item Ad-hoc tooling and manual implementation
\end{itemize} & \\ 
\end{tabular}

% \vspace{1em} % 增加垂直间距

% --- RQ3 Section ---
\begin{tabular}{p{\textwidth}}
\toprule
\textbf{RQ3: Inhibitors When Implementing Security and Privacy Practices} \\
\midrule
\end{tabular}

\begin{tabular}{@{} >{\raggedright\arraybackslash}p{0.48\textwidth} >{\raggedright\arraybackslash}p{0.48\textwidth} @{}}
\textbf{Motivational Inhibitors} & \textbf{Resources Inhibitors} \\
\begin{itemize}
    % 在这里使用 \item 添加您的条目
    \item Indifference about privacy
    \item Functional priorities over security
    \item Unclear division of responsibility
\end{itemize}
&
\begin{itemize}
    % 在这里使用 \item 添加您的条目
    \item Individual resource constrains
    \item Lack of ecosystem support
\end{itemize} \\
\textbf{Regulational Inhibitors} &  \\
\begin{itemize} 
\item Deficiencies in legal and regulatory frameworks
\end{itemize}  & \\ \bottomrule
\end{tabular}

\end{table*}

\section{Participants' Demographics}\label{app:demographics}

Table~\ref{tab:participant_demographics_tabularx} showed the demographics of our study's participants.

\begin{table*}[h]
\centering
\caption{Participants' demographics. \\ n/a = prefer not to disclose. No = no dependence on platforms}
\label{tab:participant_demographics_tabularx}
\begin{tabularx}{\textwidth}{@{}lllXXX@{}} % 使用 tabularx 环境，宽度为页面宽度
\toprule
\textbf{\#} & \textbf{Age} & \textbf{Gender} & \textbf{Application domain of the product} & \textbf{Platform used} & \textbf{Role} \\ \midrule
P1 & 18-25 & Female & General-purpose Conversational Agents & Dify & Research Assistant \\
P2 & 18-25 & Male & Education & No & Student \\
P3 & 26-35 & Female & Application Assistance & n8n & Product Manager \\
P4 & 18-25 & Female & Healthcare & Coze & Student\\
P5 & 18-25 & Male & Education & DeepSeek & Freelance (editing, self-media operation) \\
P6 & 18-25 & Male & Software Development/Developer Tools & Cursor & Student \\
P7 & 26-35 & Male & Finance & No & Student \\
P8 & 18-25 & Female & Child Education, Finance & No & Student \\
P9 & 18-25 & Female & Emotional support & No & Student \\
P10 & 26-35 & Female & Healthcare & No & Engineer \\
P11 & 18-25 & Female & General-purpose Conversational Agents & No & Researcher \\
P12 & 26-35 & Male & Automatic operations & LangChain & Researcher \\
P13 & 26-35 & Female & Productivity/Workplace & n8n & Technical support \\
P14 & 26-35 & Female & Architecture/Design automation & No & Startup CEO \\
P15 & 26-35 & Male & Research/Knowledge Work & Coze & News worker \\
P16 & 26-35 & Male & General-purpose Agents & No & Researcher \\
P17 & 36-45 & Male & Robotics, Education & Multiple Platforms & AI application development and training \\
P18 & 36-45 & Male & Social/Entertainment & No & Engineer \\
P19 & 18-25 & Male & Geolocation recommendation & No & Student \\
P20 & 26-35 & Male & Biomedical & Langchain & Researcher \\
P21 & 18-25 & Male & Real estate, Restaurants, Medical & No & Engineer \\
P22 & 26-35 & Male & Healthcare & No & Student \\
P23 & 36-45 & Male & Software Development/Developer Tools & No & Freelancer \\
P24 & 26-35 & Male & Education & Coze & Teacher \\
P25 & 36-45 & Male & Data Analysis, Self-logging Media & Midjourney, Stable Diffusion & Engineer \\
P26 & 46-55 & Male & Autonomous Systems/Transportation & Ali Cloud Bailian & Engineer \\
P27 & 36-45 & Male & Finance record, Language learning & iOS developer tools & Self-employed developer \\
P28 & 26-35 & Female & Language learning & Custom GPT & Teacher \\ \bottomrule
\end{tabularx}
\end{table*}

\clearpage
\twocolumn

\section{Interview Script}
\label{app:interview-script}

The following is the interview script. We designed the interview script to reduce researcher-induced bias. We adopted a funnel approach, moving from general open-ended questions to specific technical probes. To avoid leading participants, we refrained from introducing specific technical terms until the participants had first articulated their own perceived threats. 

\subsection*{Introduction}

Thank you for participating in this interview study. We are researching how independent AI Agent developers understand, practise, and face challenges related to privacy and security during the development process. Today's conversation will be divided into four parts: your understanding of privacy and security, your practices in actual development, the challenges you face, and your expectations and suggestions regarding platforms and regulations. We will audio-record the entire interview, and the recording will only be used for research analysis and will not be made public. If needed, a transcript of the recording can be provided to you. During the interview, you are welcome to ask questions or share any thoughts at any time. If you have no questions, I will begin recording now.

\medskip
\noindent\emph{[Start recording]}

\subsection*{Part 1: General understanding and definition}

\begin{enumerate}
  \item How do you define AI Agent?
    \begin{itemize}
      \item What are its components? How does it operate?
    \end{itemize}

  \item In the design and operation of your AI Agent, what privacy or security issues might be involved? Please provide specific examples.

  \item How do you define user privacy?
    \begin{itemize}
      \item What constitutes user privacy? To what extent should user data be utilized to ensure privacy protection?
      \item Which categories of user data are you particularly concerned about?
    \end{itemize}

  \item Mental model (the interviewer may guide the participant to draw diagrams or write explanations):
    \begin{itemize}
      \item What user data does your AI agent collect?
      \item When your AI agent interacts with users or their environment (for example, perceiving data or acquiring information), how do you understand where the data flows first and what happens to it within your agent system or the platform you use (for example, local storage, cloud, passed to an LLM, transformed)?
      \item In your understanding, when an AI agent receives input information or perceives the environment (such as text input, cameras, searches), how is this data processed?
      \item If the agent performs certain actions (such as sending emails, fetching information, controlling devices), how does this data interact with external services? Where is this data stored when the AI agent interacts with users and the environment? How does the platform handle this data?
      \item Do you think this data will be transmitted to APIs like OpenAI? Do you use APIs like OpenAI yourself?
      \item Do you think you will use external knowledge bases or RAG functionality to assist the agent? What user information might be involved?
      \item Do you think there will be calls to other third-party APIs (for example, drawing functions, email functions)? What user information might be involved?
    \end{itemize}

  \item Do you understand how underlying platforms (for example, OpenAI, a cloud provider, an agent framework) might use the data generated from agent interactions? (such as model training, system improvements, analytics)

  \item What is your understanding of an agent's memory or knowledge base? How is this information stored and accessed, and who can access it? (for example, fine-tuning data, persistent conversation history, or custom instructions)

  \item How did you establish these understandings? Where do they come from? (for example, official documents, news, training, others' experiences, or your own speculation)
\end{enumerate}

\subsection*{Part 2: Development practices and tools}

\begin{enumerate}
  \item When developing AI agents, what stages are typically involved?
  \item At which stages are privacy and security issues considered?
  \item What specific tools are used for privacy and security consideration and mitigation?
  \item Have any legal and regulatory requirements been considered?
  \item How would you communicate privacy issues to users, inform them of potential privacy risks, and what considerations would you take into account?
\end{enumerate}

\subsection*{Part 3: Probe privacy and security scenarios}

\begin{enumerate}
  \item If your AI agent performs actions (for example, sending an email, controlling a smart device, making a purchase), how do you envision the data flowing between your agent and those external services?

  \item Data leakage and misuse risks:
    \begin{itemize}
      \item Do you think your agent might engage in excessive data collection? (For instance, many agents may request extensive data permissions to provide personalized services)
      \item What kind of user profiles do you think your agent might create, and could these pose privacy risks to users? (For example, building detailed user profiles and inferring private information that users have not explicitly disclosed)
      \item Do you think your agent might experience data leaks or share data with third parties?
    \end{itemize}

  \item System security and malicious attacks:
    \begin{itemize}
      \item Do you think your agent might be vulnerable to prompt injection attacks? (For example, being manipulated through carefully crafted malicious inputs to induce or hijack its behavior)
      \item Do you think your agent might face data poisoning issues? (For instance, attackers contaminating the training data of the AI agent)
      \item Do you think your agent might be susceptible to privacy theft attacks? (For example, attackers extracting memorized sensitive information from the AI's training data)
      \item Do you think your agent might encounter issues of misuse or unauthorized access?
    \end{itemize}
\end{enumerate}

\subsection*{Part 4: Expectations}

\paragraph{From the platform's perspective}
\begin{enumerate}
  \item What specific tools, features, or guidelines do you wish AI agent development platforms (such as OpenAI or other frameworks) would provide to help you build safer and more privacy focused agents, for example built in security scanning, privacy dashboards, standardized consent authorization processes?
  \item Do you think current platforms offer sufficient support for individual developers in understanding and implementing privacy and security best practices for agents? What are the shortcomings?
  \item How can platforms better clarify the division of responsibilities between themselves and individual AI agent developers regarding privacy and security?
  \item What role do you believe platforms should play in supporting individual developers or their agents? How can this role be made most effective?
\end{enumerate}

\paragraph{From the law's perspective}
\begin{enumerate}
  \item What role do you believe legal factors can play, what are the current shortcomings, and what kind of support is needed?
\end{enumerate}

\paragraph{From the tool's perspective}
\begin{enumerate}
  \item What specific tools do you believe are needed, what roles can these tools play, and what kind of support do you need them to provide to you?
\end{enumerate}

\paragraph{From others' perspective}
\begin{enumerate}
  \item What other types of support do you believe you still need?
\end{enumerate}

\medskip
\noindent\emph{[End of interview, thank you for participation, recording ends here.]}

\end{document}